\documentstyle[11pt,aaspp4]{article}
\received{}
\accepted{}
\journalid{}{}

\slugcomment{}
\lefthead{Tsuchiya, Korchagin, Wada}
\righthead{}
\newcommand{\be}{\begin{equation}}
\newcommand{\ee}{\end{equation}}
\begin{document}

\title{Formation of Plumes in Head-on Collisions of Galaxies}
\author{Toshio Tsuchiya}
\affil{ Department of Astronomy, Faculty of Science, Kyoto University,
Kyoto 606-01,Japan\\
Email: tsuchiya@kusastro.kyoto-u.ac.jp }
\author{Vladimir Korchagin}
\affil{Institute of Physics, Stachki 194, Rostov-on-Don, Russia\\
Email: vik@rsuss1.rnd.runnet.ru}
\and
\author{Keiichi Wada$^1$}
\affil{National Astronomical Observatory, Mitaka, Tokyo 181, Japan\\
     Email: wada@th.nao.ac.jp}
\altaffiltext{1}{Space Telescope Science Institute, 3700 San Martin Drive, Baltimore, MD, 21218}
\begin{abstract}
Using N-body and SPH modeling we perform 3D numerical simulations of
head-on collisions between gas rich disk galaxies, including collisions
between counter-rotating disks and off-center collisions.  Pure stellar
intruders do not produce gaseous plumes similar to those seen in the
Cartwheel and VII Zw466 complexes of interacting galaxies; the presence
of gas in an intruder galaxy and radiative cooling are important for the
formation of a gaseous plume extending from the disk of a target galaxy.
A noticeable plume structure can be formed if the mass of an intruder is
a few percent of the mass of the primary.

The halo of the intruder is stripped in the collision, and dispersed
particles form a broad stellar bridge connecting the two galaxies. The
fraction of the intruder's halo dispersed in the collision depends on
the total mass of the intruder, and low-mass intruders lose most of
their mass.
\end{abstract}
%
%
      \section{INTRODUCTION}
%
Two decades ago, ring galaxies formed in galactic collisions were 
considered to be rare and "exotic" members of the overall galactic population.  
Now, however, the idea that
interactions and collisions might be important in galactic 
morphological transformations is
attracting more attention.
Interest in the physics of galactic
collisions was recently stimulated by the realization
that collisions and mergers were more frequent in the past
(Lavery et al. 1996),
and that collisions might be important in explaining
the diversity of forms observed in high redshift galaxies.

Lynds \& Toomre (1976) and Theys \& Spiegel (1977) developed a picture
of the ring-making process. They demonstrated that a head-on or nearly head-on
collision of a disk galaxy with an intruder generates kinematical
epicyclic oscillations excited by the gravitational field of the
intruder. The perturbed orbits of gas and stars in the galactic disk
form a fast outwardly propagating density wave concentrated in radial zones
where the stellar and/or gaseous orbits crowd together. This picture has
governed the interpretation of the
properties of ring galaxies (see e.g., Appleton \& Struck-Marcell 1996).
 
Observations provide growing evidence that the physics of galactic
collisions is more complicated and cannot be completely explained by this
simple kinematical picture. One of the long-standing problems of the
kinematical approach is the small mass of the companions in known examples
of this phenomenon.  The masses of the companions of the Cartwheel galaxy and
of VII Zw466 estimated recently by Higdon (1996) and Appleton, Charmandaris, \& Struck (1996) do not exceed six percent of the mass of the target galaxies. Such
companions cannot trigger significant density enhancements in the
disks of the primaries. This problem was first pointed out by Davies \& 
Morton (1982), who suggested that the collision may play a decisive role in triggering
a propagating star-forming wave.

Another problem was recognized by studying the neutral hydrogen
distribution in complexes of interacting ring galaxies. Higdon
(1996) and Appleton, Charmandaris, \& Struck (1996), discovered broad plumes of neutral
hydrogen connecting the Cartwheel and the VII Zw466 ring galaxies with
their companions. Our work shows that such plumes cannot be gravitationally 
dragged by a
collisionless intruder. The existence of gas plumes unambiguously demonstrate the importance of
the hydrodynamical effects in the collisions. Indeed, such effects are confirmed in
recent numerical experiments by Appleton, Charmandaris, \& Struck (1996), and Struck (1997),
who illustrated the importance of hydrodynamical effects in simulations
of colliding gaseous disks balanced by the potentials of rigid
halos.

The aim of our paper is to study the formation of plumes in galactic
collisions.  We follow the dynamics of a two-component star-gas disk
centrifugally balanced by the gravity of a rigid halo. An intruder
galaxy is modeled self-consistently, and its two-component disk is
balanced by the potential of a halo.  This model allows us to
follow the dynamics of the gas component in the colliding galaxies
simultaneously with the dynamics of the collisionless halo of the
intruder.

Recently, Struck (1997) performed extensive numerical studies of galaxy
collisions which incorporated hydrodynamics, radiative cooling, and heating by
star formation. Our study provides complementary information about
hydrodynamical effects and dynamics of the companion's halo, depending
on the mass of the companion.

We find that the formation of a ring
in the disk of the primary does not depend essentially on the admixture of gas. 
On the contrary, the presence of gas in the intruder is crucial for the
formation of plumes connecting two interacting galaxies.
An intruder with a mass
equal to ten percent of the mass of the primary does not form any noticeable ring
in its disk; the gas bridge is the main "fingerprint"
of the interaction. 

Our simulations also
demonstrate that a bridge is formed by stripping the intruder's collisionless
halo.
The fate of the halo of the intruder depends on its mass.
Most of halo particles of low-mass intruders are dispersed after the collision,
but the central cores of massive intruders can survive the
collision. In all cases, however, the collision forms a broad
plume of halo particles connecting both galaxies.

In section 2 we describe the numerical method and the model. We present 
numerical results in section 3, and in section 4 we summarize our conclusions 
and discuss the observational implications of our results. 
%
	\section{MODELS AND NUMERICAL METHOD}
%

\subsection{Models}
We examine the collisions of galaxies with disks 
containing both gas and stars, assuming that the galaxies are initially
both isolated and in equilibrium.

The equilibrium models we employ are mainly the same as those used in
the simulations of Hernquist \& Weil (1993), hereafter HW. The disks have
exponential radial profiles and vertical structures corresponding to 
isothermal sheets (Spitzer 1942). In cylindrical coordinates the
density profile of the disks is given by the expression:
\begin{equation}
\rho_d(R,z)= M_d/ ( 4\pi h^2 z_d) \exp[-R/h] {\rm sech}^2[z/z_d],
\end{equation}
where $M_d$ is the disk mass, $h$ is a radial scale length and $z_d$ is
a vertical scale thickness.

The halos of both galaxies are modeled by spherically symmetric 
distributions mimicking isothermal
potentials with a sharp cut-off at $r=r_{cut}$:   
\begin{equation}
\rho_h(r) = \left\{
\begin{array}{ll}
\rho_h(0)\left[ 1+ \left(r/r_{core} \right)^2\right]^{-1} &
 ( r \leq r_{cut}) \\
0 & ( r > r_{cut}) .
\end{array}
\right .
\end{equation}
Here $\rho_h(0)$ is the central density , $r_{core}$ is the core radius.
The mass of the halo is expressed therefore by
\begin{equation}
M_h = 4\pi \rho_h(0) r_{core}^3 \left[ \frac{r_{cut}}{r_{core}}-
\arctan\left(\frac{r_{cut}}{r_{core}}\right) \right].
\end{equation}

Our halo model differs from that of HW in the manner of truncation of
the isothermal density distribution. This, however, does not make an essential
difference in the halo dynamics if the cut-off radius is
sufficiently large. In our simulations we choose $r_{cut}=10\,r_{core}$.

The velocity distributions of the halos are isotropic Maxwellians
with velocity dispersions $\sigma_h^2$ determined by solving 
the second moment of the collisionless Boltzmann equation 
\begin{equation}
\frac{d(\rho_h \sigma_h^2)}{dr} = - \rho_h \frac{d\Phi_h}{dr}.
\end{equation}
In this equation we ignore the contribution of the disk gravity in
comparison to the gravity of the massive halo.

For the stellar disks, the velocity distribution is given  
by the velocity dispersions in the $R$, $z$ and $\phi$ directions
($\sigma_R^2$, $\sigma_z^2$, $\sigma_\phi^2$),  
and the
streaming velocity in the $\phi$ direction. The profile of
each quantity is determined by using the procedure described by Hernquist 
(1993). Determining $\sigma_R$, we set the Toomre parameter of the
axisymmetric stability $Q=1.7$  
at $R=2h$. $Q$ varies with $R$ and has a minimum at approximately $R=2h$.

For the gaseous disk we assume an isothermal equation of state with a
temperature, $T_g$. The rotational velocity of gas is determined by 
balance of gravity, pressure gradient and centrifugal
force.

\subsection{Units and Parameters of the Model}

In our simulations we employ a system of units in which the gravitational
constant $G=1$, the total disk mass of the primary $M_{d1}=1$, and the
exponential scale length of the primary stellar disk $h_{s1}=1$.  
Table 1 gives the parameters of the primary, the massive
companion, and the low-mass companion listed in these units. 


\placetable{table1}

Scaling units to values appropriate for the Milky Way parameters, namely
$M_{d1}=5.6\times10^{10} \, M_{\sun}$ and $h_{d1}=3.5$ kpc (e.g., Bahcall \&
Soneira 1980), implies the units of time and velocity are, respectively
$1.3\times10^7$ yr and 260 km s$^{-1}$. The
temperature of the isothermal gas, $T_g$, is taken to be $1.2\times10^4$ K.

\subsection{Initial Conditions}

At the beginning of simulations, the centers of
mass of the primary and the companion galaxy are placed along $z$-axis at
$z=0$ and $z=16$ respectively, with the rotation axes of both disks
pointing in $z$ direction. The companion was given the initial velocity $v_z=-1.0$
towards the primary. With these initial conditions
galaxies, accelerated by their gravity, collide
at approximately $t=12.5$.

We also simulated off-center collisions. In these simulations,
the massive companion had an initial velocity in $x$
direction $v_x=0.2$. The other parameters were the same as in the  
head-on collisions. 
With the stated value of the $x$-velocity of the primary,
the distance between the centers of both galaxies at the moment of the collision
is about two radial scale lengths appropriate to the disk of the primary.

\subsection{Numerical Method}

For all simulations presented in this work, we used {\sc treesph}
(Hernquist \& Katz 1989, hereafter HK), which computes the gravitational forces
between particles using a hierarchical tree algorithm, and calculates the
hydrodynamical interaction between gas particles using 
smoothed particle
hydrodynamics (SPH; Monaghan 1992 for review).

In the tree algorithm, we used a tolerance parameter $\theta=0.75$ and a
softening length $\epsilon=1/32$. Quadrupole terms were not included in 
the multipole expansion. In the SPH algorithm, artificial viscosity
is introduced by the HK formula, which includes bulk viscosity, with the
parameters $\alpha=0.5$ and $\beta=1.0$. We avoid solving gas energy equation
by assuming an isothermal equation of state. 

The equations of motion are integrated numerically using a leap-frog
algorithm, with individual particle time steps ranging from maximum value
$\Delta t=1/4$ 
to a minimum value $\Delta t=1/512$. The time step is
governed by the requirement that the relative energy error for a single step
is less than 0.01. For SPH particles, the time step is limited by setting the
Courant number ${\cal C}=0.3$.

The number of particles for each component is listed in Table 1.  
To save computational time, we did not solve the equations of motion for
the halo of the primary, and its distribution was fixed throughout
the simulations. We performed test simulations with self-consistent halos
and found no essential difference in the course of the evolution of the disk
of the primary or the halo of the companion.

%
	\section{RESULTS}
%
We performed simulations using 'massive' intruders with total mass equal
to twenty-five percent the total mass of the primary, and 'low mass'
intruders with mass equal to one-tenth that of the primary.
\subsection{Disk Response}

Figure \ref{fig:Mgas1-xy} shows the time evolution of the gas component
of the primary, displayed face-on to the disk plane. Before the collision
(which starts at approximately $t = 12.0$) the gaseous component develops
a transient spiral structure similar to those reported in simulations of
an isolated disk galaxy (Barnes \& Hernquist 1996). These spiral
perturbations are probably generated by the swing amplification of the
random noise in the SPH simulations, and have little influence on the
overall dynamics of gas in the colliding galaxies.


\placefigure{fig:Mgas1-xy}
\placefigure{fig:Mstr1-xy}

At later times the head-on collision develops a ring phenomenon -- this
process has been studied in detail by many authors.  Our Figure
\ref{fig:Mgas1-xy} is similar in many respects to Figure 4 of HW. The
response of the stellar disk of the primary is also similar to the
results of the previous computations.  The stellar disk
(Fig. \ref{fig:Mstr1-xy}) develops an expanding ring of enhanced density
which is broader than the ring structure in the gaseous subsystem of the
primary.


\placefigure{fig:Lgas1-xy}
\placefigure{fig:Lstr1-xy}

Our computations confirm the strong dependence of the ring amplitude
on the mass of an intruder. Figures \ref{fig:Lgas1-xy} and
\ref{fig:Lstr1-xy} show the response of the gaseous and stellar
components of the primary perturbed by a low mass intruder with mass
equal to ten percent of the mass of the primary. In agreement with the
previous studies, we did not find any noticeable ring structure generated
by the low mass intruder.


\placefigure{fig:off-center}

An off-center collision with a massive companion causes a global
redistribution of gas and stars in the primary.  The impact forms a
crescent-shaped transient structure clearly seen in the distribution of
the gas in the disk of the primary (Fig. \ref{fig:off-center}).  The well developed
crescent structure is not formed immediately after the impact of the
companion ($T=12.5$ in our run). Rather, the crescent density enhancement forms
at times $T\sim 16-22$, when the companion is already out of the disk.
 
The formation of a crescent-shaped feature by an off-center
collision has been shown previously. Appleton \&
Struck-Marcell (1987) found similar structures in a simpler
two-dimensional cloud fluid model where the companion was considered as a
perturbing force on the gaseous disk. Athanassola, Puerari, \& Bosma (1997)
found similar structures in their modeling of pure stellar disks.
We did not find any qualitative
differences between our results and the results of 
Appleton \& Struck-Marcell (1987), and more realistic properties of the
companion
do not influence the generation of crescent-shaped rings in
gravitating disks.

We note a remarkable resemblance between  
structures that appeared in the off-center collisions and the morphology
of some ring galaxies. The galaxy IIZw28 and the Wakamatsu-Nishida ring
galaxy in Sextans ( Marston \& Appleton 1995) might be good examples of
such encounters.

Well developed crescent structures do not form
immediately during the collision. In our simulations they develop  
after time $T\sim 6$, and 
this fact can be used to estimate the collisional epoch 
of off-centered ring galaxies.

\subsection{Plume Formation}
\subsubsection{Gaseous plume}
An edge-on view of the collision provides more information. Figure
\ref{fig:Mgas1-xz} shows the evolution of the gaseous components of two
colliding galaxies. The relevant feature seen in the edge-on view is a
broad gaseous plume connecting the disks of both interacting galaxies.


\placefigure{fig:Mgas1-xz}

The presence of gas in the companion is crucial for the formation of
a plume structure.
We performed simulations   
involving pure stellar intruders and did not 
find any noticeable gaseous plumes.
Similar results were found also in the numerical simulations of 
HW and Appleton \& Struck-Marcell (1996).

 
\placefigure{fig:Lgas1-xz}

The gaseous disk of the companion shown in Figure \ref{fig:Mgas1-xz} 
initially had one-fourth the gas mass of the primary. We found, however,
that a collision with a low-mass companion also forms a well-developed plume.
In the sequence shown in Figure \ref{fig:Lgas1-xz} the gas
mass of the companion is equal to ten percent the gas mass of the
primary. Nevertheless our computations demonstrate the well developed
plume connecting the disks of both galaxies.  We would like to stress that
low-mass companions do not produce ring structures in the disk of the
primary, and that plumes are the main features appearing in the collisions.

\placefigure{fig:adiabatic}

Radiative cooling of gas is also important for the formation of
plumes. For a comparison we performed simulations using an adiabatic
equation of state. Figure \ref{fig:adiabatic} illustrates results of
these simulations.  The mass of the the companion was chosen to be ten
percent of the mass of the primary, and the polytropic index has been
assumed to be equal 1.67. Contrary to the dynamics shown in Figure
\ref{fig:Lgas1-xz}, gas pushed out of the disk of the primary 
soon disperses, and does not form a plume-like structure. The
difference in the gas behavior for an adiabatic equation of state
in comparison to an isothermal one is explained by the higher velocity
dispersion gained by the adiabatic gas in the collision.

The physical mechanism of the formation of gaseous plumes due to head-on
collisions of two spiral galaxies is basically similar to the ram
pressure stripping of the interstellar matter of galaxies moving through
intergalactic gas. The approximate condition of plume formation from gas
stripped from the companion can be written as:
\be
 \Gamma_2 \equiv {R_2 \rho_{d1} V^2 \over 2 G \rho_{d2} M_2} > 1.
\label{Gamma}
\ee
Here $\rho_{d1}$ and $\rho_{d2}$ are the gas densities of the primary
and of the companion, $V$ is the relative velocity of the galaxies at
the collision, and $M_2$ and $R_2$ are the total mass and the halo radius
of the companion.
In our simulations, the relative velocity of impact of a massive
companion is $V\sim 1.5$, $M_2 \sim$ 1.0 and 0.4, $R_2 \sim$ 3.0 and
2.0, and $\rho_{d1} / \rho_{d2} \sim$ 1.0 and 2.0 for a massive and a
low-mass companion, respectively.  Substituting these values into
the expression (\ref{Gamma}) we get $\Gamma_2 \sim$ 3.4 and 11.3 for the
massive and the low-mass companion. 

As discussed in the next section,
about twenty
five percent of the massive companion, and about fifty percent of the low-mass
companion is stripped during the interaction. As a result,
the gravitational potential of an intruder becomes more shallow, and
$\Gamma_2$ is about 5 for the massive companion, and about 23 for the
low-mass one. This estimate shows that the plume formation occurs more
easily in a less massive companion, which indeed finds confirmation in
our results.

The potential of a primary is deeper than that of the companion, and
only a small fraction of primary's gas can escape from the galaxy. 
An estimate of the stripping parameter 
of the primary, $\Gamma_1 \equiv R_1 \rho_{d2} V^2/ 2 G \rho_{d1} M_1 $
gives $ \Gamma_1 \sim 1$. This estimate shows that
most of the gaseous plume is
made of the companion's gas, and our numerical results are
consistent with this argument.

The above estimate gives only a rough picture of gas
stripping, and
there is a difference between galactic collisions and stripping of  
interstellar medium of galaxies by intergalactic matter. 
Part of gas energy gained in galactic
collisions is radiated away in shocks. Simulations
show, however, that the gas gains enough kinetic energy to be
stripped.

The gaseous plume structures are relatively stable, and 
survive throughout our simulations.  In the right-bottom frame of 
Figure \ref{fig:Lgas1-xz} there is an edge-on view of the gas distribution 
when the distance between the disks of both galaxies is about a few galactic
scales. The plume structure is clearly 
seen in the simulations.
  

\placefigure{fig:Lgass-zvz}
 
The  phase-space sequence of $v_z$ vs $z$ frames shown in Figure
\ref{fig:Lgass-zvz} illustrates the dynamics of the velocity field of
the gaseous plume formed by the low mass companion.  At the beginning,
the z-components of gas velocities in both disks are equal to zero
relative to each disk. The impact starting approximately at time $t =
12.5$ forms a stream of particles with negative z-velocities ranging
from zero to the velocity of the impact of the companion. Later, the
gravity of the halo of the primary turns the low-velocity particles
towards the disk of the primary, and the velocity field of the plume can
be described as an infall onto the two galaxies.  At time $t = 64.0$,
when the separation between the disks of two galaxies is of order a few
galactic scales, approximately fifty percent of particles in the plume
have positive velocities.  Similar velocity fields were found also in
simulations of Appleton, Charmandaris, \& Struck (1996).

\subsubsection{The stellar plume}

Self-consistent modeling of the intruder's halo provides new information 
about the dynamics of the colliding galaxies.  

\placefigure{fig:Mstr2-xz}
\placefigure{fig:Lstr2-xz}

It is known that head-on collisions pump internal energy into the star
clusters (see e.g., Binney \& Tremaine 1987).  Figure
\ref{fig:Mstr2-xz} illustrates the pumping effect. It displays the
dynamics of the halo of the massive companion after the collision. A
number of particles in the companion gain kinetic energy in the
encounter.  The companion's halo expands after the collision and forms a
broad plume connecting the primary and the companion.
The behavior of the halo of a low mass companion
(Fig. \ref{fig:Lstr2-xz}) is even more dramatic.  The pumping of
internal energy into the halo of the companion causes its continual
expansion during the computation, and the volume occupied by halo
increases by two orders of magnitude.

The conclusion which can be made from our computations is that plumes
connecting two interacting galaxies contain not only gas, but also dark
matter objects and 
low-luminosity stars stripped from the halos of the companions.

The energy gain in the encounters can be estimated using the ``impulse
approximation" (Binney and Tremaine 1987). 
If the time of interaction is shorter than the orbital period of the
halo particles, then the internal energy gaining can be estimated as:
\be
  \Delta E = \pi \int_0^{\infty} [ \Delta V_R(R)]^2 \Sigma (R) R dR
{\mbox {.}}
\label{eq:deltaE}
\ee
The ratio of the pumping and binding energies for a Plummer model
can be written then as follows:
\be
{\Delta E \over E_2} \sim {GM_1 \over V^2 a_1} {M_1 \over M_2} {a_2
\over a_1}
\label{eq:relchg}
\ee
In the expressions (\ref{eq:deltaE}) and (\ref{eq:relchg}),
$M_{1\mbox{,}2}$ and $a_{1\mbox{,}2}$ are the masses and core radii of the
primary and the companion, respectively, $\Delta V_R(R)$ is the velocity
change of the particles during the collision, $\Sigma (R)$ is the
companion's density projected on the surface perpendicular to the line
of motion, and $V$ is the velocity of the encounter.

With the help of the expression (\ref{eq:relchg}) one can estimate that
the ratio of the pumped energy to the binding energy is about two for a
low-mass companion, and about unity for a massive one.  In both cases
the gain in the internal energy is considerable, and leads to
irreversible consequences for the companions.
  
Not all parts of the halo participate equally in the gain of
internal energy. The orbital periods
of particles in the outer parts of halo are an order of
magnitude longer than the time of interaction, and the "impulse" approximation
is valid for both low mass and massive companions.
The orbital periods in the central core of the halo can
be estimated as the ratio of the core size to the central velocity
dispersion.  In the core of the massive companion the orbital periods 
are comparable to the interaction time. When the orbital periods
are shorter than the interaction time, the gain of
internal energy is small, and all pumped energy is deposited in the outer
layers of the halo. Thus the outer halo is dispersed and
only the small central core survives after the collision.
This behavior is illustrated on Figures \ref{fig:Mstr2-eng} and
\ref{fig:Lstr2-eng}. They show the time dependence of the average
specific energy of eight spherical shells of the companion's halo
containing 12.5 percent each of the halo's mass.

\placefigure{fig:Mstr2-eng}
\placefigure{fig:Lstr2-eng}

Figures \ref{fig:Mstr2-eng} and \ref{fig:Lstr2-eng} illustrate clearly
that the pumped energy is mostly accumulated in the outer layers of the
companion's halo.  About thirty percent of the massive companion's halo
particles (Fig \ref{fig:Mstr2-eng}) gain enough energy after the collision
to escape the system.  These particles will eventually be
dispersed in intergalactic space and form a plume.
The influence on a low mass companion is more significant.  About
seventy percent of the halo particles gain positive specific energy and
leave the halo. The radii of six outer shells (Fig \ref{fig:Lstr2-eng})
increase with time, and only the central core of a low-mass halo stops
expanding (at approximately $t = 30$).
 
%
\section{CONCLUSIONS AND DISCUSSION}
%

In this paper we have studied the collisional interaction of disk
galaxies using SPH and N-body numerical simulations.
The specific goal of our paper was to study the formation of plumes in
head-on collisions of galaxies. Our results can be summarised as follows:\\
\begin{enumerate}
\item  We found that the formation of a ring
in the disk of a primary does not depend essentially on the admixture of
gas. However, the presence of gas in the intruder is crucial for
the formation of plumes connecting the two interacting galaxies.  A low-mass
intruder with a gas mass of about a few percent of the primary's gas
content forms a well developed plume. Most of the gaseous plume is
made of the companion's gas.

\item In agreement with previous studies, we found that the amplitude
of the outwardly propagating ring strongly depends on the relative mass
of the intruding galaxy. An intruding galaxy with a mass equal to ten
percent of the mass of the primary does not form any noticeable ring
structure in the disk of the primary, and the bridge becomes the main
"fingerprint" of the interaction.

\item  Finally, we demonstrated that a direct collision forms a
bridge of $\it particles$ stripped from the halo of the companion.
The fate of the halo of an intruder depends on its mass.  Most of
the halo particles
of a low-mass intruder are dispersed after the collision, but the
central cores of the halos of more massive intruders survive the
collision.
\end{enumerate}

Observationally, the most prominent structures of interacting galaxies
are the large-scale rings of star formation. In galaxies with low-mass
companions, these rings require some amplifying mechanism. One
possibility is the hydrodynamical effects accompanying massive star
formation. The induced nature of massive star formation might be the
missing physical factor required to explain the high efficiency of ring
formation by low-mass companions, and hence needs to be taken into account
in future work.

The broad gaseous plumes recently discovered near ring galaxies
are independent tracers of galactic interactions.  Our results indicate that
collisions produce gaseous plumes similar to those revealed in recent
observations. A comparison of the properties of the HI plumes in the
Cartwheel complex of galaxies and in VII Zw466 with the results of
our simulations supports the collisional scenario for plume formation.
The global distribution of HI in the Cartwheel complex of galaxies
reveals a massive gaseous plume with a total mass of neutral hydrogen of
about $2 \times 10^9 M_{\odot}$, which is about $16 \%$ of the HI content
of the Cartwheel, and a few times more than the total mass of the neutral
hydrogen in the presumed intruder dwarf galaxy G3 (Higdon 1996).

The gaseous plume expanding from VII Zw466 has a total mass about $5.8
\times 10^8 M_{\odot}$, which is a few times less than the mass of the
Cartwheel's plume. Such a remarkable difference in the masses of the plumes of
these two galaxies is explained by the difference in the masses of their
respective intruders. The intruder galaxy of the Cartwheel is a dwarf galaxy with
mass about $2.6 \times 10^{10} M_{\odot}$, whereas the intruder of
VII Zw466 has mass of about $7.5 \times 10^{11} M_{\odot}$, which is close
to the mass of VII Zw466 itself.  Before the collision, Cartwheel's intruder
was a gas rich dwarf galaxy. It losts most of its
gas after the collision, and the gas originally belonging to the dwarf formed
a broad and massive plume.
The intruder of the VII Zw466 galaxy had a mass comparable to the mass
of the target galaxy, and a considerable amount of gas was retained after the
collision in the deeper gravitating well of the intruder. 

One of the results of our simulations is the demonstration of the
formation of broad plumes of collisionless particles 
connecting both interacting galaxies. 
These plumes should be formed from the dark matter objects and 
the low-luminosity
stars stripped from the intruder's halo. 
If future
observations will confirm the presence of low-luminosity
stars in gaseous plumes, it will give a valuable argument 
in favor of the collisional origin of ring galaxies.

Ring galaxies can shed light on the origin of galaxies.
 According to the currently popular model of halo formation,
large galaxies like ours were built by hierarchical mergers of many
sub-galactic units (Fukugita, Hogan, \& Peebles 1996). The physics of mergers is
similar to the destruction of the companion's halo in the
direct collisions, and nearby ring galaxies provide an excellent 
opportunity to study galaxy mergers.

\begin{acknowledgments}

The authors are grateful to Greg Laughlin, Colin Norman and to 
anonymous referee 
for the careful reading of the
manuscript and suggestions which greatly improved the presentation.
TT is financially supported by Research Fellow of the Japan Society for
the Promotion of Science. VK would like to acknowledge the Yukawa
Institute for Theoretical Physics for providing financial support during
the formative phases of this project. KW thanks the Yamada Science
Foundation for the support while at STScI.

\end{acknowledgments}
\newpage
%

%
\begin{table}
\caption{Computational Parameters for our Models \label{table1}}
\begin{tabular}{llll}
\hline \hline
 & Primary & Massive companion & Low-Mass Companion \\
\hline
Halo & & & \\
mass & $ M_{h,1}=3.0 $ & $M_{h,2}=0.75$ & $M_{h,2}=0.3$ \\
core radius & $r_{core,1}=1.0$ & $r_{core,2}=0.5$ & $r_{core,2}=0.5$ \\
cut-off radius & $r_{cut,1}=10.0$ & $r_{cut,2}=5.0$ & $r_{cut,2}=5.0$ \\
particles & --- & 4096 & 4096 \\
\hline
Stellar Disk & & & \\
mass & $ M_{ds,1}=0.9 $ & $M_{ds,2}=0.225$ & $M_{ds,2}=0.09$ \\
scale length & $h_{ds,1}= 1.0$ & $h_{ds,2}= 0.5$ & $h_{ds,2}= 0.5$ \\
scale height & $z_{ds,1}= 1.0$ & $z_{ds,2}= 0.5$ & $z_{ds,2}= 0.5$ \\
$Q$ value & 1.7 & 1.7 & 1.7 \\
particles & 8192 & 4096 & 4096 \\
\hline
Gaseous Disk & & & \\
mass & $ M_{dg,1}=0.1 $ & $M_{dg,2}=0.025$ & $M_{dg,2}=0.01$ \\
scale length & $h_{dg,1}= 1.0$ & $h_{dg,2}= 0.5$ & $h_{dg,2}= 0.5$ \\
scale height & $z_{dg,1}=0.057$ & $z_{dg,2}=0.03$ & $z_{dg,2}= 0.03$ \\
$(\mbox{sound speed})^2$ & $v_{s,1}^2=10^{-3}$ & $v_{s,2}^2=10^{-3}$ &
$v_{s,2}^2=10^{-3}$ \\ 
particles & 8192 & 8192 & 8192 \\
\hline
\end{tabular}
\end{table}

\newpage
\figcaption[fig01.ps]{Time evolution of gas in the disk of
the primary seen face-on to the disk plane. Collision with a massive
companion forms an outwardly propagating ring of density enhancement
seen after T = 15.0. 
Time in all Figures is indicated at
the top right of each snapshot.
\label{fig:Mgas1-xy}}

\figcaption[fig02.ps]{Face-on view of time evolution of the stellar component
of the primary's disk in a collision with a massive companion. Collision 
forms a broad expanding ring.
\label{fig:Mstr1-xy}}

\figcaption[fig03.ps]{Evolution of gas distribution in the disk
of the primary in encounter with
a low mass companion. Collision does not form a noticeable
ring of density enhancement. 
\label{fig:Lgas1-xy}}

\figcaption[fig04.ps]{Face-on view of time evolution of the stellar
disk of the primary in encounter with a low mass companion. 
\label{fig:Lstr1-xy}}

\figcaption[fig05.ps]{A snapshot showing the distribution at T=18 of gas
in the primary in an off-center collision with a massive
companion.
\label{fig:off-center}}

\figcaption[fig06.ps]{Edge-on view of the time evolution of gas 
in both galaxies in an encounter with a massive companion. 
Collision strongly distorts the disk of the primary and forms a broad gaseous plume
connecting both galaxies.
\label{fig:Mgas1-xz}}

\figcaption[fig07.ps]{The same as on the Figure \ref{fig:Mgas1-xz}, but in encounter
with a low-mass companion. Collision forms a
broad stellar plume.
\label{fig:Lgas1-xz}}

\figcaption[figad.ps]{Time sequence of the gas evolution in interacting
galaxies with gas obeying adiabatic equation of state. The polytropic
index of gas is equal to 1.67. The total mass of the companion is
one-tenth of primary's.  Interaction of galaxies with an adiabatic
equation of state does not form a gaseous plume.
\label{fig:adiabatic}}

\figcaption[fig08.ps]{Dynamics of gas velocity field of 
the primary and the low mass companion shown at $z$ vs $v_z$ plane.  At T=0,
the $z-$ components of gas velocity in both disks are equal to zero.
After the collision the gas particles in the plume have $v_z$ - velocities
of different signs indicating that gas falls on both galaxies. 
\label{fig:Lgass-zvz}}

\figcaption[fig09.ps]{Edge-on view of time evolution of the stellar disk 
of the primary and the halo in the massive companion. Some of the halo 
particles are stripped and form a plume connecting both galaxies.
\label{fig:Mstr2-xz}}

\figcaption[fig10.ps]{The same as in the Figure \ref{fig:Mstr2-xz} but
for the encounter with a low mass companion. Most of the particles in
the low-mass halo are stripped in the collision and dispersed into
intergalactic space.
\label{fig:Lstr2-xz}}

\figcaption[fig11.ps]{Dependence on time of average radii and 
average specific energies of eight shells in the halo of the massive
companion.  Each spherical shell contains 12.5\% of the halo's mass. The
shells are labeled by ascending numbers from the center to the edge. The
three outer shells get positive specific energy after the collision and
will form a plume.
\label{fig:Mstr2-eng}}

\figcaption[fig12.ps]{The same shown on the Figure \ref{fig:Mstr2-eng}
but for the low mass companion. Most of the halo particles in the low
mass companion gain positive specific energy.
\label{fig:Lstr2-eng}}

\end{document}